\RequirePackage{ifpdf}
\ifpdf 
\documentclass[pdftex]{sigma}
\else
\documentclass{sigma}
\fi

\begin{document}

\allowdisplaybreaks

\renewcommand{\thefootnote}{$\star$}

\renewcommand{\PaperNumber}{039}

\FirstPageHeading

\ShortArticleName{Critical Exponent for Totally Asymmetric Dif\/fusion}

\ArticleName{Dynamical Critical Exponent for  Two-Species\\ Totally Asymmetric Dif\/fusion on a Ring\footnote{This paper is a contribution to the Proceedings of
the XVIIIth International Colloquium on Integrable Systems and Quantum
Symmetries (June 18--20, 2009, Prague, Czech Republic).  The full
collection is
available at
\href{http://www.emis.de/journals/SIGMA/ISQS2009.html}{http://www.emis.de/journals/SIGMA/ISQS2009.html}}}

\Author{Birgit WEHEFRITZ-KAUFMANN}

\AuthorNameForHeading{B.~Wehefritz-Kaufmann}

\Address{Department of Mathematics and Physics, Purdue University, 150 N.~University Street,\\ West Lafayette, IN 47906, USA}
\Email{\href{mailto:ebkaufma@math.purdue.edu}{ebkaufma@math.purdue.edu}}
\URLaddress{\url{http://www.math.purdue.edu/~ebkaufma/}}

\ArticleDates{Received September 28, 2009, in f\/inal form April 30, 2010;  Published online May 12, 2010}

\Abstract{We present a study of the two species totally asymmetric dif\/fusion model using the Bethe ansatz. The Hamiltonian has $U_q(SU(3))$ symmetry. We derive the nested Bethe ansatz
equations and obtain the  dynamical critical exponent from the f\/inite-size scaling properties of the
eigenvalue with the smallest real part. The dynamical critical exponent is $\frac{3}{2}$ which is the
exponent corresponding to KPZ growth in the single species asymmetric dif\/fusion model.}


\Classification{82C27; 82B20}

\renewcommand{\thefootnote}{\arabic{footnote}}
\setcounter{footnote}{0}

\section{Introduction}

The single species asymmetric dif\/fusion process has attracted and
continues to attract a lot of interest.
It is a driven dif\/fusive system describing
a stochastic movement of hard-core particles in one dimension, where
movements to the left and
right take place with dif\/ferent probabilities.
There are many interesting
applications to
shock
formation \cite{shocks,shocks2,shocks3}, traf\/f\/ic f\/low~\cite{traffic},
biopoly\-mers~\cite{biopolymers,biopolymers3,biopolymers2}
and other driven dif\/fusive systems\footnote{We are not able to give an exhaustive list of the vast literature about this subject here.}.
Since the underlying quantum spin chain is integrable, the powerful
method of
Bethe ansatz is applicable and was widely used in the past to obtain
analytical
results (see e.g.~\cite{Dhar, deGierEssler, Mallick, Gwa_Spohn, Schutz}).
The two-species asymmetric
dif\/fusion problem however  has been less investigated. While  the stationary
state has been studied in some detail~\cite{Angel, Kirone, Ferrari, Evans_Mallick, Karimipour}, much less is known about the full dynamics of the
system.

 Some results for the
 dynamical  phase diagram were obtained from the matrix product
 ansatz~\cite{Jafarpour1} (however these are restricted to certain
 regions of the full parameter space) and from numerical studies of
 the model~\cite{Godreche}.

 Recently, the multi-species asymmetric dif\/fusion
 model with open boundaries has sparked interest \cite{Ayyer, Uchiyama, Arita1}.
 This case can also be tackled using the matrix product ansatz.

 In \cite{Kim_Nijs} the dynamics of a particular choice of the model
 is studied where two species hop in opposite directions on a ring,
 the dif\/fusion constant being dif\/ferent from the passing rate of
 particles of dif\/ferent species~\cite{AHR}. This model corresponds
 directly to two coupled Burgers equations or to a Burgers equation
 coupled to a dif\/fusion equation. The
 dynamical scaling properties are calculated from the time-evolution
 of the two-point correlation functions, based on a numerical study.
 The dynamical critical exponent is consistent with a subtle  double
 Kardar--Parisi--Zhang type scaling (in the sense of factorization).

 An analytical study of the dynamical scaling of the multi-species asymmetric
exclusion process was described by \cite{Arita} who found a dynamical
critical exponent of $\frac{3}{2}$ from a Bethe ansatz calculation if the dif\/fusion
was asymmetric and a critical exponent of $2$ for symmetric dif\/fusion.

 In the present paper, we  investigate the dynamical critical exponent for the {\em  totally asymmetric} dif\/fusion process  with equal rates which is a related but dynamically dif\/ferent model that will be precisely
 def\/ined in Section~\ref{section2}.
We will derive the
Bethe ansatz equations starting from the $R$-matrix of the totally asymmetric dif\/fusion model,
calculate the energy gap and from a f\/inite-size scaling analysis
extract the  critical exponent.

\subsection{Discussion of the literature about the two-species\\ asymmetric dif\/fusion model}

We will brief\/ly survey the literature on the two-species asymmetric dif\/fusion model. Although
there seem to be several approaches building on previous work \cite{Arita, Dahmen, Popkov, Cantini} by trying to extract an appropriate limit to totally asymmetric dif\/fusion, the details of each such strategy
are not obvious, however,  and one runs into serious dif\/f\/iculties and subtleties as we will explain.
We found that instead of trying to tweak the calculations in an ad-hoc manner, it is clearer and simpler to start from scratch and derive the Bethe ansatz equations directly for the totally asymmetric case. With hindsight, our calculation provides a f\/irm ground for future comparison which might also facilitate to formulate the correct limiting procedure.

The Bethe ansatz equations for the two-species asymmetric dif\/fusion model were put forth in an article by Dahmen
\cite{Dahmen}. They
depend on a parameter $\gamma$ that is related to the asymmetric hopping rates by $\exp(\gamma)=q=\sqrt{\frac{\Gamma_L}{\Gamma_R}}$, where $\Gamma_L$ and $\Gamma_R$ describe the hopping rate to the left and right, respectively. It is not at all clear how to take the limit of one of these hopping rates going to zero directly in the Bethe ansatz equations. Since  in the limit $q\rightarrow 0$ all the ratios tend to one, and even the usual rescaling of the roots by a factor of~$q$ does not work~-- the resulting equations either diverge or become trivial.

In the paper by Popkov at al.~\cite{Popkov} Bethe ansatz equations for three-species models with dif\/ferent hopping rates for all six possible particles exchanges are derived from the matrix product ansatz. One of the cases they discuss is a particular asymmetric exclusion process where the hopping rates are coupled as $g_{B0}=g_{A0}+g_{AB}$ where $g_{B0}$, $g_{A0}$ and $g_{AB}$
correspond to the processes $A+0 \rightarrow 0+A$, $B+0 \rightarrow 0+B $ and $A+B \rightarrow B+A$, respectively (and the three other rates are set to zero). In this paper we will choose
all three non-zero rates to be the same, so the model considered here is dif\/ferent.

Although the way Cantini \cite{Cantini} derives the Bethe ansatz equations for the case of  asymmetric dif\/fusion seems to be the most amenable for taking the limit to our model, however, he
has introduced phase factors $e^{\nu_{10}}$, $ e^{\nu_{20}}$, $e^{\nu_{12}}$ that remain in the Bethe ansatz equations, and it is not clear  which value they should take for the case of totally asymmetric dif\/fusion.

In the article by Arita et al.~\cite{Arita}  the Bethe ansatz equations for the multi-species asymmetric dif\/fusion model were derived. However, the Bethe ansatz equations for the case of totally
asymmetric dif\/fusion are impossible to obtain from this calculation in a straightforward manner. Although it is possible to take the limit where one of the hopping rates is set equal to zero at the starting point, the $R$-matrix, in the subsequent calculations the authors often multiply by or divide by the parameter that would become zero in the totally asymmetric limit. So taking the limit in the f\/inal equation does not lead to
reasonable equations, and would have to do a~step by step analysis of the derivation and make the necessary alterations at each step
to avoid division by or multiplication by zero. At this point, we found it simpler and less ambiguous
to just  derive the equations for our case directly.

\section{Two-species totally asymmetric dif\/fusion}\label{section2}

\subsection{Asymmetric dif\/fusion}
The two-species asymmetric dif\/fusion model consists of two species
of particles, $A$ and $B$,  dif\/fusing asymmetrically in one dimension. The following processes
take place:
\begin{alignat*}{3}
& A+0 \rightarrow 0+A \ \mbox{with rate} \ \Gamma_R,\qquad
&& 0+A \rightarrow A+0 \ \mbox{with rate} \ \Gamma_L, &  \\
& B+0 \rightarrow 0+B \ \mbox{with rate} \ \Gamma_R,\qquad
&& 0+B \rightarrow B+0 \ \mbox{with rate} \ \Gamma_L, &  \nonumber\\
& A+B \rightarrow B+A \ \mbox{with rate} \ \Gamma_R, \qquad
&& B+A \rightarrow A+B \ \mbox{with rate} \ \Gamma_L.
\end{alignat*}

\subsection{Totally asymmetric dif\/fusion}

In the totally asymmetric case, $\Gamma_L=0$, so particles $A$ do
not dif\/fuse to the left since the interchange
$B+A \rightarrow A+B$ is blocked. This leads to a dif\/ferent dynamics for the model
that will be analyzed in this article.

\begin{figure}[t]
\centering
\includegraphics[width=0.35\textwidth]{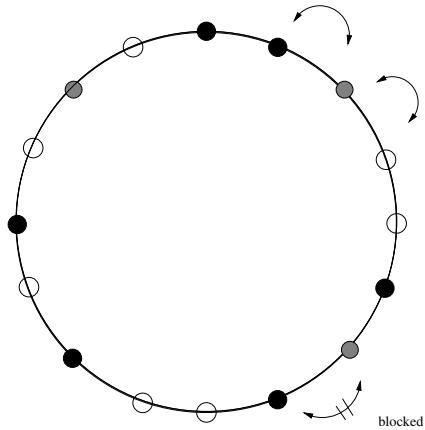}
\caption{Example of totally asymmetric dif\/fusion on a ring. Black dots correspond to $A$ particles, grey dots to $B$ particles and open dots to vacancies.}\label{Fig1}
\end{figure}

Assuming periodic boundary conditions,  i.e.\ on the ring ${\mathbb
Z}/L
{\mathbb Z}$ (where $L$ is the number of discrete sites of that
ring), the model is shown qualitatively in Fig.~\ref{Fig1}. Dif\/fusion to the right corresponds to
clockwise motion around the circle, dif\/fusion to the left to counterclockwise motion. Black dots represent particles $A$, grey dots particles $B$ and open dots vacancies. The arrows indicate
processes that are still allowed of $\Gamma_L=0$, the blocked arrow indicates the process $B+A \rightarrow A+B$ that is forbidden in the totally asymmetric case.

Particles $A$ do not see any dif\/ference between vacant sites and
particles~$B$, whereas particles~$B$ trying to dif\/fuse to the right are blocked by particles~$A$.
Therefore,
we call particles~$A$ ``f\/irst-class'', and particles~$B$
``second-class''
particles. There is an interesting connection between second-class particles and the study of current f\/luctuations~\cite{secondclass}.

\subsection{Master equation and Hamiltonian}

The dynamics of the asymmetric dif\/fusion model is described by a master equation for the pro\-ba\-bility
distribution $p_t(\eta)$ of the lattice conf\/iguration $\eta(t)$ at time $t$.
If we denote a conf\/iguration by $\eta$ and the jump rates $\Gamma_R$ and $\Gamma_L$ by $c(j,j+1,\eta)$, when interchanging the particles and/or vacancies of the conf\/iguration $\eta$  on sites $j$ and $j+1$, the master equation reads:
\begin{gather*} \frac{d}{dt} p_t(\eta)=\sum_j \left[c(j,j+1,\eta^{j j+1})p_t(\eta^{j
j+1})-c(j,j+1,\eta) p_t(\eta)\right],
\end{gather*}
where
$\eta^{j j+1}$ denotes the conf\/iguration obtained from $\eta$ by interchanging the particles and/or vacancies at sites $j$ and $j+1$.
Following~\cite{Alcaraz_Annals, Alcaraz_Rittenberg}, attaching a vector space $\mathbb{C}^3$ at each discrete point $j$
and using a vector $(1,0,0)^{T}$ for a particle of type $A$,
a vector $(0,1,0)^{T}$ for a particle of type $B$
and a vector $(0,0,1)^{T}$ for a vacancy, the
operator of the master equation can be written as a quantum spin chain which is given by the
following expression if one assumes periodic boundary conditions
\begin{gather}
H=D \sum_{j=1}^{L}  \left[\frac{q+q^{-1}}{2}-q\sum_{\alpha < \beta}
(E_j^{\alpha \beta} E_{j+1}^{\beta
\alpha})-q^{-1} \sum_{\alpha > \beta}
(E_j^{\alpha \beta} E_{j+1}^{\beta
\alpha})-\frac{q+q^{-1}}{2}\sum_{\alpha=1}^{3}(E_j^{\alpha \alpha}
E_{j+1}^{\alpha \alpha})\right.\nonumber\\
  \left. \phantom{H=}{}-\frac{q-q^{-1}}{2} \sum_{\alpha\neq \beta}
(\mbox{sign}(\alpha-\beta) E_j^{\alpha \alpha} E_{j+1}^{\beta
\beta})\right].\label{asymmetricH}
\end{gather}
In the following, the standard vector space $\mathbb{R}^3$ will just be called $V$.
In this expression, the matrices $E_j^{\alpha \beta}$ are $3 \times 3$ matrices with
only one non-zero entry: $(E_j^{\alpha \beta})_{\gamma \delta}=\delta_{\alpha \gamma} \delta_{\beta \delta}$
and, as usual, the expression $\sum\limits_{j=1}^{L-1} E_j^{\alpha \beta} E_{j+1}^{\beta
\alpha}$ means the $L$-fold tensor product
\[
 1 \!\! 1_1 \otimes 1 \!\! 1_2 \otimes \cdots
 \otimes 1 \!\! 1_{j-1} \otimes E_j^{\alpha \beta} \otimes
 E_{j+1}^{\beta \alpha} \otimes 1
 \!\! 1_{j+2}\otimes \cdots \otimes 1 \!\! 1_L.
 \]
 The parameters $D$ and $q$ are real and depend on the dif\/fusion
 rates: $q=\sqrt{\frac{\Gamma_R}{\Gamma_L}}$ and $D=\sqrt{\Gamma_R
 \Gamma_L}$.

This Hamiltonian is integrable and the eigenvalues can be found by applying the Bethe ansatz.
Since the Hamiltonian is non-Hermitian in general, we will encounter complex eigen\-va\-lues.

\subsection{Dynamical critical exponent}

In non-equilibrium dynamics, the dynamical critical exponent describes a relation between the relaxation time towards equilibrium
$\tau$ (or temporal correlation length) of a system and the spatial correlation length
 $\xi$, namely that $\tau \simeq \xi^z$ with the dynamical critical exponent $z$. It can be shown that for one-dimensional quantum spin chains, $\tau \simeq L^z$. Since the relaxation time is dominated by the eigenvalue of the Hamiltonian with the smallest real part (the energy of the ground state being equal to zero), we can obtain the exponent $z$ from a f\/inite size analysis
 of the Hamiltonian of the system as
 \begin{gather}
 \mbox{Re} \, (E_1) = \mbox{const} \, \frac{1}{L^z}. \label{scaling}
 \end{gather}.

 For the single species asymmetric dif\/fusion model, $z=\frac{3}{2}$ was obtained in~\cite{Gwa_Spohn} from a Bethe ansatz calculation for the totally asymmetric case which therefore belongs to the KPZ {\cite{KPZ}} universality class, whereas $z=2$ for the partially asymmetric case which describes the Edwards--Wilkinson universality class~\cite{EW}. We will determine the exponent~$z$ from a careful study of (\ref{scaling}). We will f\/ind the lowest lying eigenvalue of the totally asymmetric dif\/fusion model by means of the Bethe ansatz.

\subsection{Nested Bethe ansatz}

We start with the totally asymmetric dif\/fusion model, setting the parameters
$\Gamma_L=0$ and $\Gamma_R=1$. This does not lead to any singularities in the Hamiltonian since its expression contains
the products $ D q= \Gamma_R=1$ and $D q^{-1}= \Gamma_L=0$. The new Hamiltonian reads
\begin{gather}
\label{totallyasymmetricH}
H= \sum_{j=1}^{L}  \left[\frac{1}{2}-\sum_{\alpha < \beta}
(E_j^{\alpha \beta} E_{j+1}^{\beta
\alpha})-\frac{1}{2}\sum_{\alpha=1}^{3}(E_j^{\alpha \alpha}
E_{j+1}^{\alpha \alpha})\right.  \left. -\frac{1}{2} \sum_{\alpha \neq  \beta}
(\mbox{sign}(\alpha-\beta) E_j^{\alpha \alpha} E_{j+1}^{\beta
\beta})\right].
\end{gather}

This Hamiltonian is  integrable, and we will use the  algebraic Bethe ansatz to f\/ind its spectrum. Although the Bethe ansatz is well known for the Hamiltonian (\ref{asymmetricH}), the totally asymmetric case (given by the Hamiltonian (\ref{totallyasymmetricH})) cannot be obtained as a special case since setting $\Gamma_L=0$ causes the Bethe ansatz equations to become singular.
Therefore, we will derive the Bethe ansatz equations for the totally asymmetric case
following \cite{Babelon, Alacaraz_notes}. We use the following $R$-matrix elements
\begin{gather*}
R^{\alpha \alpha}_{\alpha \alpha} = \exp{\theta}\quad \mbox{for} \ \alpha=1,\ldots,3,\nonumber\\
R^{\alpha \beta}_{\beta \alpha} = 2 \sinh{\theta}\quad \mbox{for}\  \alpha <  \beta, \ \alpha, \beta =1 \ldots,3,\nonumber\\
R^{\alpha \beta}_{\beta \alpha} = 0\quad \mbox{for}\  \alpha >  \beta, \ \alpha, \beta =1, \ldots,3,\nonumber\\
R^{\alpha \beta}_{\alpha \beta}  =  \exp{\theta}\quad \mbox{for} \  \alpha <  \beta, \ \alpha, \beta =1, \ldots,3,\nonumber\\
R^{\alpha \beta}_{\alpha \beta}  =  \exp({-\theta})\quad \mbox{for}\  \alpha >  \beta, \  \alpha, \beta =1, \ldots,3,
\end{gather*}
where the indices denote the following tensor product in ${\rm End}(V \otimes V)$:
\[
R_{il}^{mk} E_{mi} \otimes E_{kl},
\] or in the language of the associated vertex model
this corresponds to the initial state denoted by the two indices $mi$ scattering into the f\/inal state
denoted by the indices $kl$.

This $R$-matrix satisf\/ies the factorization equation
\[
R^{jk}_{pq}(\theta_2-\theta_3) R^{ip}_{lr}(\theta_1-\theta_3)R^{rq}_{mn}(\theta_1-\theta_2)=
R^{ij}_{qr}(\theta_1-\theta_2) R^{rk}_{pn}(\theta_1-\Theta_3) R^{pq}_{lm}(\theta_2-\theta_3).
\]
We now def\/ine the $3^L \times 3^L$ matrices $T(\theta)^{[L]}_{ab}$ as
\[
T^{[L]}(\theta)_{ab}=\sum_{a_1,\ldots, a_{L-1}=1}^3  t_{a a_1}(\theta) \otimes t_{a_1 a_2}(\theta)\otimes \cdots \otimes t_{a_{L-1}b}(\theta),
\]
 where
\[
[t_{ab}(\theta)]_{ij} = R^{i b}_{a j}(\theta),\qquad a,b,i,j = 1,2,3,
\]
 are $3 \times 3$ matrices.
 We def\/ine the monodromy matrix as
\begin{gather}
 T^{[L]}=\left(  \begin{array}{ccc}
 T_{11}^{[L]} & T_{21}^{[L]} & T_{31}^{[L]}\\
  T_{12}^{[L]} & T_{22}^{[L]} & T_{32}^{[L]}\\
  T_{13}^{[L]} & T_{23}^{[L]} & T_{33}^{[L]}
 \end{array} \right)=\left( \begin{array}{ccc} A & B_1 & B_2\\
 C_1 & D_{11} & D_{12}\\
 C_2 & D_{21} & D_{22}
 \end{array} \right).
 \label{monodromy}
\end{gather}
 The monodromy matrix satisf\/ies the fundamental relation
 \begin{gather}
 R(\theta - \theta^{\prime}) [T^{[L]}(\theta) \otimes T^{[L]}(\theta^{\prime})]=[T^{[L]}(\theta^{\prime} ) \otimes  T^{[L]}(\theta)] R(\theta-\theta^{\prime}).
\label{RTT}
\end{gather}

The transfer matrix
\[
\tau(\theta)= \sum_{i=1}^3 T_{ii}(\theta)
\]
is obtained as the trace of the monodromy matrix. It is a matrix acting on $V^{\otimes L}$.
In the above notation, it can be written as
\[ \tau(\theta)= A (\theta)+D_{11}(\theta)+D_{22}({\theta}).
\]
We recover the Hamiltonian (\ref{totallyasymmetricH}) as the logarithmic derivative of the transfer matrix, if the spectral parameter $\theta$ is set to zero:
\begin{gather}
H = \frac{d (\log \tau(\theta))}{d \theta}\Big|_{\theta=0}.
\label{H}
\end{gather}
Therefore, diagonalizing $H$ will be the same problem as diagonalizing $\tau$.

\subsection{Diagonalization of the transfer matrix}

We start be deriving algebraic relations for the matrices $A$, $B_j$, $C_j$, $j=1,2$ and $D_{ik}$, $i,k=1,2$ in
equation~(\ref{monodromy}).

Since we would like to diagonalize $\tau$, we are looking for a reference state that is
a simultaneous eigenstate of $A$ and $D_{ii}$ and is annihilated by $C_i$ and $D_{ij}$
for $i\neq j$.

We choose
\begin{gather}|\Omega\rangle=\left(\begin{array}{ccc}1\\0\\0\end{array}\right)^{(1)}\otimes
\left(\begin{array}{ccc}1\\0\\0\end{array}\right)^{(2)}\otimes\cdots\otimes
\left(\begin{array}{ccc}1\\0\\0\end{array}\right)^{(L)}=
\otimes_{i=1}^L\left(\begin{array}{ccc}1\\0\\0\end{array}\right)^{(i)}
\label{omega}
\end{gather}
to be our reference state. This is not necessarily the physical ground state.
$T^{[L]}$ acting on this reference state is interpreted as follows:
\begin{gather}
T^{[L]}|\Omega\rangle=\left(  \begin{array}{ccc}
 T_{11}^{[L]}|\Omega\rangle & T_{21}^{[L]} |\Omega\rangle& T_{31}^{[L]}|\Omega\rangle\\
  T_{12}^{[L]}|\Omega\rangle & T_{22}^{[L]}|\Omega\rangle & T_{32}^{[L]}|\Omega\rangle\\
  T_{13}^{[L]}|\Omega\rangle & T_{23}^{[L]}|\Omega\rangle & T_{33}^{[L]}|\Omega\rangle
 \end{array} \right)=\left( \begin{array}{ccc} a(\theta)^L |\Omega\rangle & B_1|\Omega\rangle & B_2|\Omega\rangle\\
0 & c(\theta)^L |\Omega\rangle& 0\\
0 & 0 & c(\theta)^L|\Omega\rangle\\
 \end{array} \right),
 \label{T}
 \end{gather}
where $a(\theta)=\exp({\theta})$ and $c(\theta)=2 \sinh(\theta)$.
The operators $B_i(\theta,{\alpha})$
will create new states when acting on the reference state. They will be called creation operators
in the following.

In the next step we use the fundamental relation equation~(\ref{RTT}) to derive the bilinear algebra among the operators $A(\theta)$, $B_i(\theta)$, $C_{i}(\theta)$ and $D_{ij}(\theta)$.

The relations we are going to use later on are:
\begin{gather}
A(\theta) B_i(\theta^{\prime})  =  g(\theta^{\prime}-\theta) B_i(\theta^{\prime} )A(\theta)-h(\theta^{\prime}-\theta) B_i(\theta) A(\theta^{\prime}), \nonumber\\
B_i(\theta) B_j(\theta)  =  r^{ij}_{pq} B_p(\theta^{\prime})B_q(\theta),\nonumber\\
D_{ij}(\theta) B_k(\theta^{\prime})  =  g(\theta-\theta^{\prime}) r^{ik}_{pq} B_p(\theta^{\prime})
D_{jq}(\theta)-g(\theta-\theta^{\prime})B_i(\theta) D_{jk}(\theta^{\prime}).
\label{commutators}
\end{gather}
Here $r^{ij}_{pq}$ are coef\/f\/icients of a $4 \times 4 $ matrix where only the following f\/ive
entries are dif\/ferent from zero:
$ r_{11}^{11}=1$, $r^{12}_{12}=1$, $r^{21}_{12}=2 \sinh(\theta)e^{-\theta}$, $r_{21}^{21}=e^{-2 \theta}$, $r^{22}_{22}=1$ and the functions $g(\theta)$ and $h(\theta)$ are given by:
\[
g(\theta)=\frac {e^{\theta}}{2 \sinh(\theta)},\qquad h(\theta)=2 e^{-\theta}{ \sinh(\theta)}=\frac{1}{g(\theta)}.
\]
Now we are ready to look for an eigenstate of the transfer matrix $\tau(\theta)$.

We will make the following ansatz for an eigenfunction of the transfer matrix having the creation operators $B_i(\theta)$, $i=1,2$ act on the reference state $\Omega$ given by equation~(\ref{omega}).
\[
\Psi(\lambda_1,\lambda_2,\ldots,\lambda_p)=\sum_{\{\sigma\}} x_{\sigma} B_{\sigma(y_1)}(\lambda_1) B_{\sigma(y_2)} (\lambda_2) \cdots B_{\sigma(y_p)}(\lambda_p)|\Omega\rangle,
\]
where the sum runs over all possible permutations of the set of indices $\{y_1,y_2,\ldots, y_p\}
\in \{1,2\}$.

We will calculate the action of the transfer matrix $\tau$ on the vector $\Psi$.
Recall that $\tau=A(\theta)+D_{11}(\theta)+D_{22}(\theta)$. So we will encounter terms of the form
\begin{gather*}
A(\theta) B_{\sigma(y_1)}(\lambda_1) B_{\sigma(y_2)} (\lambda_2) \cdots B_{\sigma(y_p)}(\lambda_p)|\Omega\rangle
\qquad \mbox{and} \\
D_{ii}(\theta) B_{\sigma(y_1)}(\lambda_1) B_{\sigma(y_2)} (\lambda_2) \cdots B_{\sigma(y_p)}(\lambda_p)|\Omega\rangle.
 \end{gather*}
 We know how $A(\theta)$ and $D_{ii}(\theta)$ commute with the operators $B_i(\lambda_j)$ from
the relations (\ref{commutators}) and we know how $A(\theta)$ and $D_{ii}(\theta)$ act
on $\Omega$ from equation~(\ref{T}).

We will get two types of terms. In the f\/irst type of terms, we get the original
combination of $B_{\sigma(y_1)}(\lambda_1) B_{\sigma(y_2)} (\lambda_2) \cdots B_{\sigma(y_p)}(\lambda_p)|\Omega\rangle$ back. They originate from taking the f\/irst term in the commutation relation for $B_i(\theta) B_j(\theta)$ in~(\ref{commutators}). Using standard terminology, we will call those terms {\it wanted terms}. In the second type
of terms, one of the $B_{\sigma(y_i)}$ operators will depend on the parameter $\theta$. These terms will be called {\it unwanted terms} and their sum has to be set to zero.

We will also use the following notation: We def\/ine a $2^p$ dimensional vector $\mathbb{B}$ containing all possible
terms of the form $B_{\sigma(y_1)}(\lambda_1) B_{\sigma(y_2)} (\lambda_2) \cdots B_{\sigma(y_p)}$ where, as above, $\sigma$ denotes a permutation of the set of indices $\{y_1,y_2,\ldots, y_p\}
\in \{1,2\}$. We also def\/ine a $2^p$ dimensional vector $X$ containing the corresponding $x_{\sigma}$ in the same order. This allows us to rewrite the sum as a scalar product of these two vectors:
\[
\sum_{\{\sigma\}} x_{\sigma} B_{\sigma(y_1)}(\lambda_1) B_{\sigma(y_2)} (\lambda_2) \cdots B_{\sigma(y_p)}(\lambda_p) =B(\lambda_1)\otimes B(\lambda_2) \otimes \cdots  \otimes B(\lambda_p)\cdot X= \mathbb{B}\cdot X.
\]

Then the action of the transfer matrix $\tau(\theta)$ on $\Psi(\lambda_1,\lambda_2,\ldots,\lambda_p)$ leads to wanted terms that can be written as:
\[
a(\theta)^L\prod_{i=1}^p g(\lambda_i-\theta) \Psi(\lambda_1,\lambda_2,\ldots,\lambda_p).
\]

The unwanted terms arise from taking the second term in the commutation relation for the $B(\theta)$ operators (\ref{commutators}). E.g.~one of the unwanted terms appears if the second term in the commutator is applied to $A(\theta) B(\lambda_k)$ and is of the form
\[
 h(\lambda_k-\theta) a(\lambda_k)^L\prod_{n \neq k}^p g(\lambda_n-\lambda_k).
\]

As observed by de Vega \cite{deVega} the general unwanted term can easily written down if one uses the following fact about a cyclic permutation of $p$ operators $B(\lambda_j)$:
\[
B(\lambda_1)\otimes B(\lambda_2)\otimes \cdots \otimes B(\lambda_p)=B(\lambda_2) \otimes B(\lambda_3)\otimes \cdots \otimes B(\lambda_p) \otimes B(\lambda_1) \tau_{(2)}^p(\lambda_1,{ \{\lambda\}}),
\]
where $\tau_{(2)}^p(\theta,\{ \lambda\})=\sum\limits_{a=1}^2 T_{aa}^{[p]}(\theta,\{ \lambda\})$
and
\[
T_{ab}^{[p]}(\theta,\{ \lambda\})=
\sum_{a_1,\ldots, a_{p-1}=1}^2  t^{[2]}_{a a_1}(\theta-\lambda_1) \otimes t^{[2]}_{a_1 a_2}(\theta-\lambda_2)\otimes \cdots \otimes t^{[2]}_{a_{p-1}b}(\theta-\lambda_p),
\]
 where
 $[t^{[2]}_{ab}(\theta)]_{ij}$, $a,b,i,j = 1,2 $
 are $2 \times 2$ matrices where only the following f\/ive entries are non-zero:
 \begin{gather*}
 [t^{[2]}_{11}(\theta)]_{11} = r^{11}_{11}(\theta),\qquad
 [  t^{[2]}_{11} (\theta) ] _{22}   = r^{21}_{12}(\theta),\qquad
  [t^{[2]}_{21}(\theta)]_{21}  =  r^{21}_{21}(\theta),\\
 [t^{[2]}_{12}(\theta)]_{12}   = r^{12}_{12}(\theta), \qquad
 [t^{[2]}_{22}(\theta)] _{22}  =  r^{22}_{22}(\theta).
\end{gather*}
Here $r^{ib}_{aj}(\theta) $ are the coef\/f\/icients appearing in equation~(\ref{commutators}) and
$\tau_{(2)}^p(\theta,\{ \lambda\})$ is the transfer matrix of the six-vertex model (i.e.\ a model with two states instead of three states) for a line of $p$ sites with inhomogeneities $\{\lambda_i\}$, $i=1, \ldots,p$.

So the cyclic permutation $B(\lambda_i) \rightarrow B(\lambda_{i+1})$ followed by the multiplication of the matrix
\[
M= \tau_{(2)}^p(\lambda_1,\{ \lambda\})
\]
leaves the product $B(\lambda_1)\otimes B(\lambda_2) \otimes \cdots \otimes  B(\lambda_p)$ invariant.
Therefore the general unwanted term in the result for $A(\theta) \Psi(\lambda_1 \cdots \lambda_p)$ can be written as
\begin{gather*}
-\sum_{k=1}^{p} \{h(\lambda_k-\theta) a(\lambda_k)^L \prod_{n=1,n \neq k}^p g(\lambda_n-\lambda_k) B(\theta) \otimes B(\lambda_{k+1}) \otimes \cdots\nonumber \\
\qquad{} \otimes B(\lambda_p) \otimes B(\lambda_1) \otimes \cdots \otimes B(\lambda_{k-1}) M^{k-1} \}X |\Omega\rangle.
\end{gather*}
Similarly one can f\/ind the wanted and unwanted terms after acting with $(D_{11}(\theta)+D_{22}(\theta))$ on~$|\Psi\rangle$. The wanted term is
\begin{gather*}
(D_{11}(\theta)+D_{22}(\theta)) |\Psi\rangle= \prod_{j=1}^{p} g(\theta-\lambda_j) c(\theta)^L B(\lambda_1) \otimes B(\lambda_2) \otimes\cdots \otimes B(\lambda_p) \tau_{(2)}^{(p)}(\theta,\{\lambda\}) X |\Omega\rangle.
\end{gather*}
Using the same argument as before, the general unwanted term can be written as a sum of terms where $B(\theta)$ replaces $B(\lambda_k)$:
\begin{gather*}
\sum_{k=1}^p\{ -h(\theta-\lambda_k) \prod_{n-=1,n \neq k}^p g(\lambda_k-\lambda_n)c(\lambda_k)^L   B(\theta)\otimes B(\lambda_{k+1})\otimes\cdots\\
\qquad{}\otimes B(\lambda_p) \otimes B(\lambda_1)\otimes \cdots \otimes B(\lambda_{k-1})M^{k-1}
\tau_{(2)}^{(p)}(\lambda_k,\{\lambda\})\} X |\Omega\rangle.
\end{gather*}
Altogether, the sum of wanted terms reads
\begin{gather*}
B(\lambda_1) \otimes B(\lambda_2) \cdots \otimes B(\lambda_p)[ a(\theta)^L \prod_{i=1}^p g(\lambda_i-\theta)+
\prod_{j=1}^{p} g(\theta-\lambda_j)   c(\theta)^L\tau_{(2)}^{(p)}(\theta,\{\lambda\}) ]X |\Omega\rangle.
\end{gather*}
The sum of unwanted terms reads:
\begin{gather*}
-\sum_{k=1}^{p}  [B(\theta) \otimes B(\lambda_{k+1}) \otimes \cdots \otimes B(\lambda_p) \otimes B(\lambda_1) \otimes \cdots \otimes B(\lambda_{k-1})]M^{k-1}  \\
\qquad{}\times
\Bigg\{ h(\lambda_k-\theta) a(\lambda_k)^L\prod\limits_{n=1,n \neq k}^p g(\lambda_n-\lambda_k) + h(\theta-\lambda_k) \prod_{n-=1,n \neq k}^p g(\lambda_k-\lambda_n) \\
\qquad{}\times c(\lambda_k)^L
\tau_{(2)}^{(p)}(\lambda_k,\{\lambda\}) \Bigg\}X |\Omega\rangle.
\end{gather*}
Since we are looking for an eigenstate of the transfer matrix, the sum of all wanted terms has to be proportional to $|\Psi\rangle$ and the sum of all unwanted terms has to be equal to zero.
The f\/irst condition determines $X$ to be an eigenvector of $\tau_{(2)}^{(p)}$ corresponding to the eigenvalue $\Lambda_{(2)}(\theta,\{\lambda\})$:
\begin{gather}
\tau_{(2)}^{(p)} X=\Lambda_{(2)}(\theta,\{\lambda\}) X.
\label{eigenvalueeq}
\end{gather}

The eigenvalue $\Lambda_{(2)}(\theta,\{\lambda\})$ is determined by requiring the sum of unwanted terms to become zero:
\begin{gather}
\Lambda_{(2)}(\lambda_k,\{\lambda\})=\left(\frac{a(\lambda_k)}{c(\lambda_k)}\right)^L \prod_{n=1,n\neq k}^p \frac{g(\lambda_n-\lambda_k)}{g(\lambda_k-\lambda_n)}.
\label{lambda}
\end{gather}
We have reduced
 the original eigenvalue problem to
equation~(\ref{eigenvalueeq}) which is an eigenvalue equation for the transfer matrix $\tau_{(2)}^{(p)}$ of the six-vertex model. This is
the crucial step in this paper. All we have to do now is repeat the same diagonalization procedure for equation~(\ref{eigenvalueeq}). We repeat exactly the same steps as before, acting with the transfer matrix on a reference state with $r$ creation matrices B, and deriving equations for the wanted and unwanted terms.

This leads to another expression for the eigenvalue $\lambda_{(2)}$
that can be set equal to equation~(\ref{lambda}) and another consistency equation stemming from setting the unwanted terms equal to zero. In this way we obtain
the following nested Bethe ansatz equations
which have two types of
unknowns, $\lambda_k$, $k=1,\ldots,p$ and $\Lambda_j$, $j=1,\ldots,r$,
\begin{gather}
\left[\frac{\exp(\lambda_k)}{2
\sinh(\lambda_k)}\right]^L = \prod_{\stackrel{s=1}{s\neq
k}}^{p}(-\exp{(2
\lambda_k- 2 \lambda_s)})\prod_{j=1}^r \frac{\exp(\Lambda_j-\lambda_k)}{2
\sinh(\Lambda_j-\lambda_k)},\qquad k=1,\dots,p,\nonumber\\
\prod_{k=1}^{p}\frac{\exp{(\Lambda_j-\lambda_k)}}{2
\sinh(\Lambda_j-\lambda_k)} = \prod_{\stackrel{n=1}{n\neq
j}}^r(-\exp{(2 \Lambda_j-2 \lambda_n)}),\qquad j=1,\ldots,r.\label{BA}
\end{gather}

The eigenvalues of $H$  are obtained using the relation between the transfer matrix and the Hamiltonian given by equation~(\ref{H}):
\begin{gather}
\label{BAenergy}
E=\frac{d}{d \theta} \ln( \Lambda_{(2)}(\theta))\big|_{\theta=0}=L+\sum_{k=1}^p \frac{\exp{\lambda_k}}{\sinh(\lambda_k)}.
\end{gather}

\subsection{Numerical solution of the Bethe ansatz equations}

Since the Hamiltonian (\ref{totallyasymmetricH}) is describing a reaction-dif\/fusion system, the ground state is
the steady state with energy zero and all excitation energies will have positive real parts.
We f\/ixed~$L$,~$p$ and~$r$ and solved the coupled system of Bethe ansatz equations
numerically. We also diagonalized the Hamiltonian (\ref{totallyasymmetricH})
numerically for a small number of sites (typically up to $L=9$). We compared the
energies obtained from the Bethe ansatz with the ones obtained by
numerical diagonalization of Hamiltonian $H$ to single out the
eigenvalue with the smallest non-zero real part which plays the role of the second lowest eigenvalue and determines the gap. This f\/irst excited state  always lies in the sector
with $p=\frac{L}{3}$, $r=0$. This state has an equal density of particles $A$, $B$ and vacancies of $\frac{1}{3}$. We  extrapolated the energy values of the f\/irst excited state for
$L\rightarrow \infty$.

\subsection{Results}

\begin{figure}[t]
\centering
\includegraphics[width=.6\textwidth]{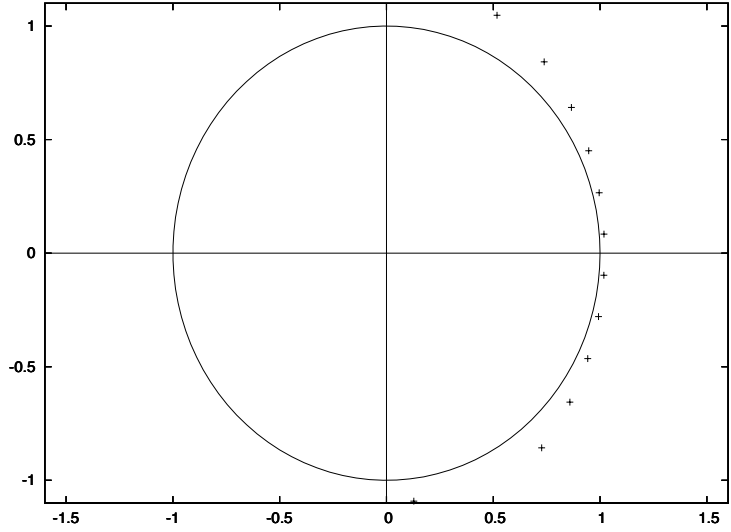}
\caption{Complex roots $\lambda_k$, $k=1,\ldots,12$ for the Bethe ansatz equations with $p=\frac{L}{3}$, $r=0$ and $L=36$.}\label{Fig2}
\end{figure}

We found the solution for $p=\frac{L}{3}$, $r=0$  corresponding to the
f\/irst excited state of Hamiltonian. To make sure that this really is the lowest lying excited state, we compared to the results of a~numerical diagonalization of the Hamiltonian for f\/inite lattice lengths. This state has an equal density of particles $A$, $B$ and vacancies of $\frac{1}{3}$. The roots all lie in the complex plane. A typical pattern of roots for
the f\/irst excited state of the Hamiltonian is  shown for $L=36$ in Fig.~\ref{Fig2}. The solid line is the unit circle that is given as a guide to the eye.

The data was extrapolated using the Bulirsch--Stoer algorithm \cite{BST}. Since we expect a scaling of the form (remember the ground state energy is always zero)
\[
{\rm Re}\, (\Delta E(L))= {\rm Re}\,(E_1(L))= \mbox{const}\, L^{-z} +o(L^{-z}),
\] we built extrapolants for the exponent $-z$ as
\[
\frac{{\rm Log}\left(\frac{{\rm Re}\,(\Delta E(L))}{{\rm Re}\,(\Delta E(L+3))}\right)}{{\rm Log}\left(\frac{L}{L+3}\right)}.
\]

The data used in the extrapolation for the exponent is shown in table:
$$\mbox{\begin{tabular}{cc}
     $L$  & \mbox{extrapolant}\\
     \hphantom{0}6  &  $-1.6336892192762$\\
     \hphantom{0}9  &  $-1.6252314332778$\\
     12  &  $-1.6092183117219$\\
     15  &  $-1.5952666540982$\\
     18  &  $-1.5839870664789$\\
     21  &  $-1.5749003909369$\\
     24  &  $-1.5674968193872$\\
     27  &  $-1.5613778750522$\\
     30  &  $-1.5562495252464$\\
     33  &  $-1.5518961566109$
\end{tabular}}
$$

\noindent
The result of the extrapolation with BST-algorithm is
	 $-z=-1.50000009$ with error $0.00000323$. This clearly shows that the exponent $z$
	 is $\frac{3}{2}$.

	 It is interesting to note that this state corresponds to the Bethe ansatz equations with only one type of roots.  The vanishing of the second type of roots does not correspond to the simple inclusion of the one-species model into the two-species model given by making the density of the second type of particles zero but rather for the state we consider corresponds to equal densities of these in-equivalent particles. The fact that the densities are coupled may suggest that there is some underlying quasi-particle formalism which may give a theoretical way to explain the connection to the single species exclusion model and the occurrence of the exponent $\frac{3}{2}$ well known from the totally asymmetric single species exclusion model.

\subsection{Analytical solution of the Bethe ansatz equations}

First we would like to rewrite the Bethe ansatz equations equations (\ref{BA}) and (\ref{BAenergy})
in integral form. To that end we use the following two changes of variables:
$e^{\lambda_k}=z_k$, $z_k^2 =Z_k$, $k=1, \ldots, p$ and $e^{\Lambda_j}=y_j$, $y_j^2=Y_j$, $j=1,\ldots, r$.
Equation (\ref{BA}) becomes:
\begin{gather}
\left( \frac{Z_k}{Z_k -1}\right)^L  =
\prod_{\stackrel{s=1}{s\neq
k}}^{p}\left(-\frac{Z_k}{Z_s}\right)\prod_{j=1}^r \frac{Y_j}{(Y_j-Z_k)},\qquad k=1,\dots,p, \nonumber\\
\prod_{k=1}^{p}\left(\frac{Y_j}{Y_j-Z_k}\right) = \prod_{\stackrel{n=1}{n\neq
j}}^r\left(-\frac{Y_j}{Y_n}\right), \qquad j=1,\ldots,r. \label{newBA}
\end{gather}

The new equation for the energies reads:
\begin{gather*}
E=L+\sum_{k=1}^p \frac{2 Z_k}{Z_k-1}.
\end{gather*}

\begin{figure}[t]
\centering
\includegraphics[width=.6\textwidth]{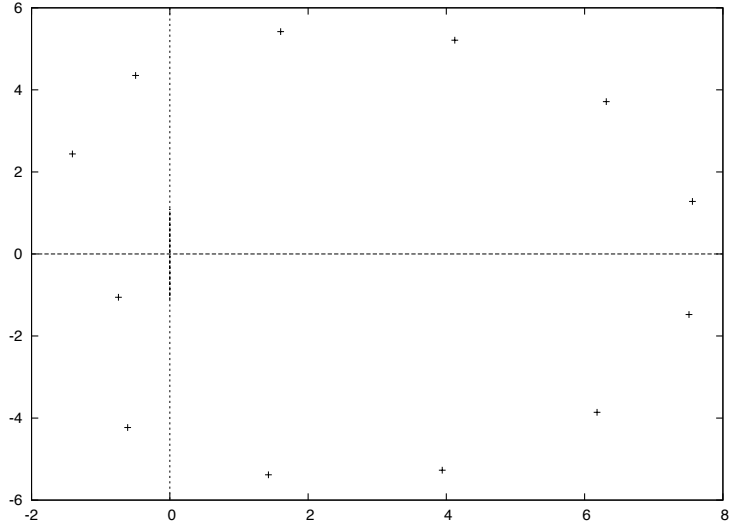}
\caption{Transformed complex roots $Z_k$, $k=1,\ldots,12$ for the Bethe ansatz equations with $p=\frac{L}{3}$, $r=0$ and $L=36$.}\label{Fig3}
\end{figure}

Our numerical work described above suggests that the f\/irst excited state lies in the sector with $p=\frac{L}{3}$, $r=0$.  After applying the above transformations, the transformed roots $Z_k$, $k=1, \ldots, \frac{L}{3}$ will lie on a curve that is shown for $L=36$ in Fig.~\ref{Fig3}.

In order to analyze the Bethe ansatz equations in the limit of large lattice lengths $L$, it is convenient to introduce
a function
\begin{gather*}
g(z)=\ln\left(\frac{z}{z-1}\right)
\end{gather*}
and a function
\[
K(z_l,z)=\ln\left(\frac{z}{z_l}\right).
\]
For both def\/initions the branch cut of the ln-function is taken along the negative real axis.
The Bethe ansatz equations for  $p=\frac{L}{3}$, $r=0$ can now be written as
\[
Y_L(Z_j)=\frac{2 \pi}{L} I_j,\qquad j=1,\ldots,\frac{L}{3}
\]
with a so-called counting function
\[
iY_L(z)=g(z)+\frac{1}{L} \sum_{l=1}^{L/3} K(z_l,z).
\]

Each excited state of the Hamiltonian in the sector with $p=\frac{L}{3}$, $r=0$   will correspond to one set of integers $\{I_j\,|\,j=1,\ldots,\frac{L}{3}\}$.

The situation is very similar to the one described in \cite{deGierEssler2} for the one-species asymmetric exclusion model. Therefore we can adopt the technique to transform the discrete Bethe ansatz equations (\ref{newBA}) to integral equations by using an identity that follows from the residue theorem:
\begin{gather*}
\frac{1}{L}\sum_{j=1}^{{L/3}} f(z_j)=\frac{1}{4 \pi i}\oint_C dz f(z)Y_L^{\prime} (z) \cot \left (\frac{1}{2} L Y_L(z)\right ).
\end{gather*}

\begin{figure}[t]
\centering
\includegraphics[width=.5\textwidth]{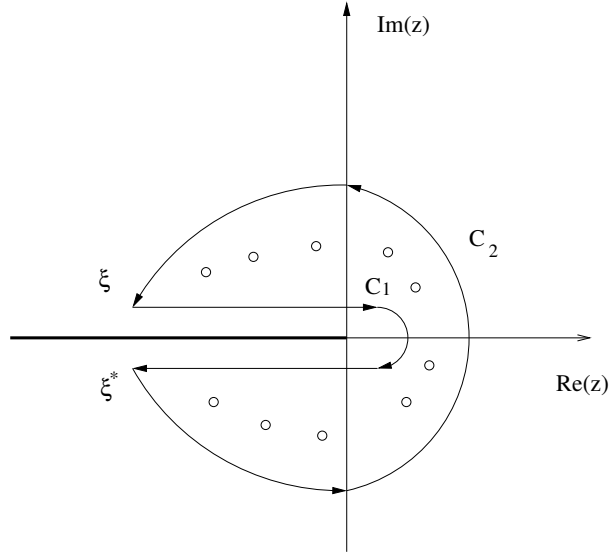}
\caption{Sketch of the integration contour. The open dots represent the roots $Z_j$.}\label{Fig4}
\end{figure}

In this equation, $C$ is the contour enclosing all the roots $Z_j$. We can view this contour $C$ as the union of two contours $C_1$ and $C_2$ as shown in Fig.~\ref{Fig4}.

These two contours intersect at two points $\xi$ and $\xi^{\star}$. We will f\/ix those two points by requiring
\[
Y_L(\xi^{\star})=-\pi+\frac{\pi}{L}, \qquad Y_L(\xi)=\pi-\frac{\pi}{L}.
\]

Rewriting this integral by separating the contributions coming from two contours
$C_1$ and $C_2$, this becomes:
\begin{gather*}
i Y_L(z)=g(z)+\frac{1}{2 \pi}\! \int_{\xi^{\star}}^{\xi}\! K(w,z)
Y_L^{\prime}(w) dw+\frac{1}{2 \pi} \! \int_{C_1} \!\frac{K(w,z)
Y_L^{\prime}}{1{-}e^{-iLY_L(w)} }dw
+\frac{1}{2 \pi} \!\int_{C_2}\! \frac{K(w,z) Y_L^{\prime}}{e^{iLY_L(w)}{-}1 }dw.
\end{gather*}
The formula for the energy reads:
\[
E=L+\frac{L}{2 \pi} \int_{\xi^{\star}}^{\xi} \epsilon(z)
Y_L^{\prime}(w) dw+\frac{L}{2 \pi} \int_{C_1} \frac{\epsilon(z)
Y_L^{\prime}}{1-e^{-iLY_L(w)} }dw
+\frac{L}{2 \pi} \int_{C_2} \frac{\epsilon(z) Y_L^{\prime}}{e^{iLY_L(w)}-1 }dw,
\]
where the function $\epsilon(z)$ is given by
\[
\epsilon(z)=\frac{2 z}{z-1}.
\]

It should be possible to analyze these equations along the lines of de Gier and Essler \cite{deGierEssler2} by an expansion in inverse powers of $L$; this will be left for a future publication.

\section{Conclusion and outlook}
We have shown that the dynamical critical exponent  of the totally
asymmetric exclusion model is~$\frac{3}{2}$ which in the single-species asymmetric dif\/fusion model is the exponent for the KPZ universality class. This is a new result that
cannot be deducted from the recent paper by Arita et al.~\cite{Arita}.

The Bethe ansatz equations and their solutions are qualitatively very
dif\/ferent from the case of a single species asymmetric exclusion
model. It would be very
interesting to derive this result analytically once the patterns of
the solutions of the Bethe ansatz equations are fully understood.
Building on the results reported here, the next step would be to solve the integral equations representing the Bethe ansatz equations in the continuum
along the lines of de~Gier and Essler~\cite{deGierEssler2}. Since the integration is performed over the curve formed by the roots in the thermodynamic limit, a qualitative understanding of this curve is a prerequisite for this calculation.
We hope to report on this soon.

Another new direction will be to generalize the Bethe ansatz to the case of open boundaries.

\subsection*{Acknowledgements}

We would like to thank V.~Rittenberg for his continued interest and invaluable discussions and F.C.~Alcaraz for sharing his manuscript about the Bethe ansatz with us. We would also like to acknowledge support from the Purdue Research Foundation.

\pdfbookmark[1]{References}{ref}
\LastPageEnding


\begin{thebibliography}{99}

\footnotesize\itemsep=0pt

\bibitem{shocks}
Ferrari P.A.,
Shocks in one-dimensional processes with a drift,
 in Probability and Phase Transition (Cambridge, UK, July 4--16, 1993), Editor G.~Grimmett,
{\it NATO ASI Ser., Ser. C, Math. Phys. Sci.}, Vol.~420,
 Kluwer, Dordrecht, 1994, 35--48.\\
Ferrari P.A., Fontes L.R.G.,
Shock f\/luctuations in the asymmetric simple exclusion process,
\href{http://dx.doi.org/10.1007/BF01199027}{{\it Probab. Theory Related Fields}} {\bf 99} (1994), 305--319.

\bibitem{shocks2}
Derrida B., Lebowitz J.L., Speer E.R.,
Shock prof\/iles for the asymmetric simple exclusion process in one dimension,
\href{http://dx.doi.org/10.1007/BF02770758}{{\it J. Stat. Phys.}} {\bf  89} (1997), 135--167,
\href{http://www.arxiv.org/abs/cond-mat/9708051}{cond-mat/9708051}.\\
 Bal\'azs M.,
 Microscopic shape of shocks in a domain growth model,
\href{http://dx.doi.org/10.1023/A:1012271624597}{{\it J. Stat. Phys.}} {\bf  105}  (2001), 511--524,
\href{http://www.arxiv.org/abs/math.PR/0101124}{math.PR/0101124}. \\
 R\'akos A.,  Sch\"{u}tz G.M.,
 Exact shock measures and steady-state selection in a driven dif\/fusive system with two conserved densities,
\href{http://dx.doi.org/10.1023/B:JOSS.0000044064.62295.29}{{\it  J. Stat. Phys.}} {\bf 117} (2004), 55--76,
\href{http://www.arxiv.org/abs/cond-mat/0401461}{cond-mat/0401461}.\\
B\'alazs M., Farkas G., Kovacs P.,  R\'akos A.,
Random walk of second class particles in product shock measures,
\href{http://dx.doi.org/10.1007/s10955-010-9933-8}{{\it J. Stat. Phys.}} {\bf 139} (2010), 252--279,
\href{http://www.arxiv.org/abs/0909.3071}{arXiv:0909.3071}.

 \bibitem{shocks3}
 Jafarpour F.H.,
 Multiple shocks in a driven dif\/fusive system with two species of particles,
 {\it Phys.~A}  {\bf 358} (2005), 413--422,
 \href{http://www.arxiv.org/abs/cond-mat/0504093}{cond-mat/0504093}.\\
 Jafarpour F.H., Masharian S.R.,
The study of shocks in three-states driven-diffusive systems: a matrix product approach,
\href{http://dx.doi.org/10.1088/1742-5468/2007/03/P03009}{{\it J. Stat. Mech. Theory Exp.}} {\bf 2007}  (2007), P03009, 18~pages, \href{http://www.arxiv.org/abs/cond-mat/0612622}{cond-mat/0612622}.


\bibitem{traffic}
Chowdhury A., Santen L., Schadschneider A.,
Statistical physics of vehicular traf\/f\/ic and some related systems,
\href{http://dx.doi.org/10.1016/S0370-1573(99)00117-9}{{\it Phys. Rep.}} {\bf  329} (2000), 199--329,
\href{http://www.arxiv.org/abs/cond-mat/0007053}{cond-mat/0007053}.

\bibitem{biopolymers}
Macdonald J.T., Gibbs J.H., Pipkin A.C.,
Kinetics of biopolymerization on nucleic acid templates,
\href{http://dx.doi.org/10.1002/bip.1968.360060102}{{\it Biopolymers}} {\bf  6} (1968), 1--25.
	
 \bibitem{biopolymers3}
 Sch\"{u}tz G.,
 Non-equilibrium relaxation law for entangled polymers,
\href{http://dx.doi.org/10.1209/epl/i1999-00529-8}{{\it Europhys. Lett.}} {\bf  48} (1999), 623--628.

\bibitem{biopolymers2}
 Widom B., Viovy J.L., Defontaines A.D.,
 Repton model of gel-electrophoresis and dif\/fusion,
\href{http://dx.doi.org/10.1051/jp1:1991239}{{\it J. Phys. I France}} {\bf  1} (1991), 1759--1784.


\bibitem{Dhar}
Dhar D.,
An exactly solved model for interfacial growth,
{\it Phase Transitions} {\bf  9} (1987), 51--86.

\bibitem{deGierEssler}
de Gier  J., Essler F.,
Exact spectral gaps of the asymmetric exclusion process with open boundaries,
\href{http://dx.doi.org/10.1088/1742-5468/2006/12/P12011}{{\it J. Stat. Mech. Theory Exp.}} {\bf 2006} (2006), P12011, 46~pages,
\href{http://www.arxiv.org/abs/cond-mat/0609645}{cond-mat/0609645}.

\bibitem{Mallick}
Golinelli  O., Mallick K.,
Derivation of a matrix product representation for the asymmetric exclusion process from algebraic Bethe ansatz,
\href{http://dx.doi.org/10.1088/0305-4470/39/34/004}{{\it J. Phys. A: Math. Gen.}} {\bf  39} (2006), 10647--10658,
\href{http://www.arxiv.org/abs/cond-mat/0604338}{cond-mat/0604338}.

\bibitem{Gwa_Spohn}
 Gwa L.H., Spohn H.,
 Bethe solution for the dynamical-scaling exponent of the noisy Burgers equation,
 \href{http://dx.doi.org/10.1103/PhysRevA.46.844}{{\it Phys. Rev. A}}  {\bf 46} (1992), 844--854.

\bibitem{Schutz}
Sch\"{u}tz G.M.,
Exact solution of the master equation for the asymmetric exclusion process,
\href{http://dx.doi.org/10.1007/BF02508478}{{\it J. Stat. Phys.}}  {\bf 88} (1997), 427--445,
\href{http://www.arxiv.org/abs/cond-mat/9701019}{cond-mat/9701019}.

\bibitem{Angel}
Angel O.,
The stationary measure of a 2-type totally asymmetric exclusion process,
\href{http://dx.doi.org/10.1016/j.jcta.2005.05.004}{{\it J. Combin. Theory Ser.~A}} {\bf  113} (2006), 625--635,
\href{http://www.arxiv.org/abs/math.PR/0501005}{math.PR/0501005}.

\bibitem{Kirone}
Evans M.R., Ferrari P.A., Mallick K.,
Matrix representation of the stationary mesure for the multispecies TASEP,
\href{http://dx.doi.org/10.1007/s10955-009-9696-2}{{\it J. Stat. Phys.}}  {\bf 135} (2009), 217--239,
\href{http://www.arxiv.org/abs/0807.0327}{arXiv:0807.0327}.



\bibitem {Ferrari}
Ferrari P.A.,  Martin J.B.,
Stationary distributions of multi-type totally asymmetric exclusion processes,
\href{http://dx.doi.org/10.1214/009117906000000944}{{\it Ann. Probab.}} {\bf 35} (2007), 807--832,
\href{http://www.arxiv.org/abs/math.PR/0501291}{math.PR/0501291}.


\bibitem{Evans_Mallick}
 Prolhac S., Evans  M.R.,  Mallick K.,
 The matrix product solution of the multispecies partially asymmetric exclusion process,
\href{http://dx.doi.org/10.1088/1751-8113/42/16/165004}{{\it J. Phys. A: Math. Theor.}}  {\bf 42} (2009), 165004, 25~pages,
\href{http://www.arxiv.org/abs/0812.3293}{arXiv:0812.3293}.


\bibitem{Karimipour}
Karimipour V.,
A multi-species asymmetric exclusion process, steady state and correlation functions on a~periodic lattice,
\href{http://dx.doi.org/10.1209/epl/i1999-00389-2}{{\it Europhys. Lett.}} {\bf 47} (1999), 304--310,
\href{http://www.arxiv.org/abs/cond-mat/9809193}{cond-mat/9809193}.


\bibitem{Jafarpour1}
Jafarpour F.H.,
A two-species exclusion model with open boundaries: a use of $q$-deformed algebra,
\mbox{\href{http://www.arxiv.org/abs/cond-mat/0004357}{cond-mat/0004357}}.


\bibitem{Godreche}
Evans M.R., Foster D.P., Godreche C., Mukamel D.,
Asymmetric exclusion model with two species: spontaneous symmetry breaking,
\href{http://dx.doi.org/10.1007/BF02178354}{{\it J. Stat. Phys.}} {\bf 80} (1995), 69--102. \\
 Evans M.R., Foster D.P., Godreche C., Mukamel D.,
 Spontaneous symmetry breaking in one dimensional driven dif\/fusive system,
\href{http://dx.doi.org/10.1103/PhysRevLett.74.208}{{\it Phys. Rev. Lett.}} {\bf  74} (1995), 208--211.

\bibitem{Ayyer}
Ayyer A., Lebowitz J., Speer E.R.,
On the two species asymmetric exclusion process with semi-permeable boundaries,
\href{http://dx.doi.org/10.1007/s10955-009-9724-2}{{\it J. Stat. Phys.}}  {\bf 135} (2009), 1009--1037,
\href{http://www.arxiv.org/abs/0807.2423}{arXiv:0807.2423}.




\bibitem{Uchiyama}
Uchiyama M.,
Two-species asymmetric exclusion process with open boundaries,
\href{http://dx.doi.org/10.1016/j.chaos.2006.05.013}{{\it Chaos Solitons Fractals}} {\bf 35} (2008), 398--407,
\href{http://www.arxiv.org/abs/cond-mat/0703660}{cond-mat/0703660}.


\bibitem{Arita1}
Arita C.,
Phase transitions in the two-species totally asymmetric exclusion process with boundaries,
\href{http://dx.doi.org/10.1088/1742-5468/2006/12/P12008}{{\it J. Stat. Mech. Theory Exp.}} {\bf 2006} (2006), P12008, 19~pages.


\bibitem{Kim_Nijs}
 Kim K.H., den Nijs M.,
 Dynamic screening in a two-species asymmetric exclusion process,
\href{http://dx.doi.org/10.1103/PhysRevE.76.021107}{{\it Phys. Rev. E}} {\bf 76} (2007), 021107, 14~pages,
\href{http://www.arxiv.org/abs/0705.1377}{arXiv:0705.1377}.


\bibitem{AHR}
Arndt P.F., Heinzel T., Rittenberg V.,
Spontaneous breaking of translational invariance and spatial condensation in stationary states on a ring.
I.~The neutral system,
\href{http://dx.doi.org/10.1023/A:1004670916674}{{\it J. Stat. Phys.}} {\bf  97} (1999), 1--65,
\href{http://www.arxiv.org/abs/cond-mat/9809123}{cond-mat/9809123}.


\bibitem{Arita}
Arita C., Kuniba A., Sakai K., Sawabe T.,
Spectrum in multi-species  simple exclusion process on a ring,
\href{http://dx.doi.org/10.1088/1751-8113/42/34/345002}{{\it J.~Phys.~A: Math. Theor.}} {\bf 42} (2009), 345002, 41~pages,
\href{http://www.arxiv.org/abs/0904.1481}{arXiv:0904.1481}.

\bibitem{Dahmen}
Dahmen S.R.,
Reaction-dif\/fusion processes described by three-state quantum chains and integrability,
\href{http://stacks.iop.org/0305-4470/28/905}{{\it J.~Phys.~A: Math. Gen.}} {\bf 28} (1995), 905--922,
\href{http://www.arxiv.org/abs/cond-mat/9405031}{cond-mat/9405031}.

\bibitem{Popkov}
Popkov V., Fouladvand M.E., Sch\"{u}tz G.M.,
 A suf\/f\/icient criterion for integrability of stochastic many-body dynamics and quantum spin chains,
\href{http://dx.doi.org/10.1088/0305-4470/35/33/314}{{\it J. Phys. A: Math. Gen.}} {\bf 35} (2002), 7187--7204,
\href{http://www.arxiv.org/abs/hep-th/0205169}{hep-th/0205169}.

\bibitem{Cantini}
Cantini L.,
Algebraic Bethe ansatz for the two-species ASEP with dif\/ferent hopping rates,
\href{http://dx.doi.org/10.1088/1751-8113/41/9/095001}{{\it J. Phys. A: Math. Theor.}} {\bf 41} (2008), 095001, 16~pages,
\href{http://www.arxiv.org/abs/0710.4083}{arXiv:0710.4083}.

\bibitem{secondclass}
 Pr\"{a}hofer M., Spohn H.,
 Current f\/luctuations for the totally asymmetric simple exclusion process,
 in In and Out of Equilibrium (Mambucaba, 2000), Editor V.~Sidoravicius, {\it Progr. Probab.}, Vol.~51, Birkh\"auser Boston, Boston, MA, 2002,  185--204,
 \href{http://www.arxiv.org/abs/cond-mat/0101200}{cond-mat/0101200}.\\
 B\'alazs M., Sepp\"{a}l\"{a}inen T.,
 Exact connections between current f\/luctuations and the second class particle in a class of deposition models,
 \href{http://dx.doi.org/10.1007/s10955-007-9291-3}{{\it J. Stat. Phys.}} {\bf 127} (2007), 431--455,
 \href{http://www.arxiv.org/abs/math.PR/0608437}{math.PR/0608437}.



\bibitem{Alcaraz_Annals}
Alcaraz F.C., Droz M., Henkel M., Rittenberg, V.,
Reaction-dif\/fusion processes, critical dynamics and quantum chains,
\href{http://dx.doi.org/10.1006/aphy.1994.1026}{{\it Ann. Physics}}  {\bf 230} (1994), 250--302,
\href{http://www.arxiv.org/abs/hep-th/9302112}{hep-th/9302112}.

\bibitem{Alcaraz_Rittenberg}
Alcaraz F.C., Rittenberg, V.,
Reaction-dif\/fusion processes as physical realizations of Hecke algebras,
\href{http://dx.doi.org/10.1016/0370-2693(93)91252-I}{{\it Phys. Lett. B}} {\bf 314} (1993), 377--380,
\href{http://www.arxiv.org/abs/hep-th/9306116}{hep-th/9306116}.



\bibitem{KPZ}
Kardar M., Parisi G., Zhang Y.C.,
Dynamic scaling of growing interfaces,
\href{http://dx.doi.org/10.1103/PhysRevLett.56.889}{{\it Phys. Rev. Lett.}} {\bf  56} (1986), 889--892.

\bibitem{EW}
 Wilkinson D.R., Edwards S.F.,
The surface statistics of a granular aggregate,
\href{http://dx.doi.org/10.1098/rspa.1982.0057}{{\it Proc. Roy. Soc. London Ser.~A}} {\bf 381} (1982), no.~1780, 33--51.

\bibitem{Babelon}
 Babelon O., de Vega H.J., Viallet C.-M.,
Exact solution of the $Z_{n+1} \times Z_{n+1}$ symmetric generalization of the $XXZ$ model,
\href{http://dx.doi.org/10.1016/0550-3213(82)90087-6}{{\it Nuclear Phys.~B}}  {\bf 200} (1982),  266--280.

\bibitem{Alacaraz_notes}
Alcaraz C.,
Lecture notes, unpublished.

\bibitem{deVega}
 de Vega H.J.,
 Yang--Baxter algebras, integrable theories and quantum groups,
\href{http://www.ams.org/leavingmsn?url=http://dx.doi.org/10.1142/S0217751X89000959}{{\it Internat. J. Modern Phys. A}} {\bf 4} (1989), 2371--2463.



\bibitem{BST}
  Bulirsch R., Stoer J.,
 Fehlerabsch{\"{a}}tzungen und Extrapolation mit rationalen Funktionen bei Verfahren vom Richardson-Typus,
\href{http://dx.doi.org/10.1007/BF01386092}{{\it Numer. Math.}} {\bf 6} (1964), 413--427.\\
 Christe  P., Henkel M., Introduction to conformal invariance and its applications to critical phenomena, {\it Lecture Notes in Physics, New Series m: Monographs}, Vol.~16, Springer-Verlag, Berlin, 1993.

\bibitem{deGierEssler2}
de Gier  J., Essler F.H.L.,
Slowest relaxation mode of the partially asymmetric exclusion process with open boundaries,
\href{http://dx.doi.org/10.1088/1751-8113/41/48/485002}{{\it J. Phys. A: Math. Theor.}} {\bf 41}  (2008), 485002, 25~pages,
\href{http://www.arxiv.org/abs/0806.3493}{arXiv:0806.3493}.



\end{thebibliography}
\end{document}